\documentclass[fleqn,12pt,twoside]{article}
\usepackage[headings]{espcrc1}

\readRCS
$Id: espcrc1.tex,v 1.2 2004/02/24 11:22:11 spepping Exp $
\usepackage{graphicx}
\usepackage[figuresright]{rotating}

\newcommand{\AmS}{{\protect\the\textfont2
  A\kern-.1667em\lower.5ex\hbox{M}\kern-.125emS}}

\title{Strangeness Production at SIS measured with HADES}

\author{J. Pietraszko\address[MCSD]{GSI Helmholtz Centre for Heavy Ion Research
GmbH Planckstrasse 1, \\D-64291 Darmstadt, GERMANY}%
        \thanks{e-mail: j.pietraszko@gsi.de}
        L. Fabbietti\address{Excellence Cluster 'Universe', Technische Universit\"{a}t M\"{u}nchen,
Boltzmannstr. 2, D-85748 Garching Germany.}  (for the HADES
collaboration)}

\begin{document}

\maketitle

\begin{abstract}
In this paper we review the recent results on strangeness
production measured by HADES in the Ar+KCl system at a beam energy
of 1.756 AGeV. A detailed comparison of the measured hadron yields
with the statistical model is also discussed.
\end{abstract}

\section{Hadron production in Ar+KCl collisions}
The study of nuclear matter properties at high densities and
temperatures is one of the main objectives in relativistic
heavy-ion physics. In this context, the paramount aim of measuring
the particle yields emitted from heavy-ion collisions is to learn
about the K-N potential, production mechanism of strange particles
or the nuclear equation of state. Strangeness production in
relativistic heavy-ion collisions at SIS/Bevelac energy range has
been extensively studied by various groups including experiments
at the Bevelac\cite{bevelac_1} and at SIS, KaoS\cite{kaos_1},
FOPI\cite{fopi_1}, and in recent years also HADES.\\ Recently, the
HADES\cite{hades_nim} collaboration measured charged particle
production in the Ar+KCl system\cite{hades_kaons} at 1.756 AGeV.
Although HADES was designed primarily for di-electron
measurements, it has also shown an excellent capability for the
identification of a wide range of hadrons like $K^-$, $K^+$,
K$^0$$\rightarrow$$\pi^+\pi^-$, $\Lambda$$\rightarrow$p$\pi^-$,
$\phi$$\rightarrow$$K$$^+$$K$$^-$ and even
$\Xi$$\rightarrow$$\Lambda$$\pi^-$. The measured yields and slopes
of the transverse mass of kaons and $\Lambda$ particles have been
found to be in good agreement with the results obtained by
KaoS\cite{kaos_0} and FOPI\cite{fopi_0}. The dependencies of the
measured $K^-/K^+$ ratio on the centrality and on the collision
energy follow the systematics measured by KaoS \cite{hades_kaons}
too.\\A combined and inclusive identification of $K^+K^-$ and
$\phi$ mesons was performed for the first time in the same
experimental setup at subthreshold beam energy for $K^-$ and
$\phi$ production. The obtained $\phi/K^-$ ratio of
0.37$\pm$0.13$\%$ indicates that 18$\pm$7$\%$ of kaons stem from
$\phi$ decays. Since the $\phi$ mesons reconstructed via $K^+K^-$
channel are those coming mainly from decays happening outside the
nuclear medium, this value should be considered as a lower limit.
In addition the non-resonant $K^+K^-$ production can contribute to
the measured $K^-$ yield. Unfortunately, this part is not known in
heavy-ion collisions, but it has been measured to be about 50$\%$
of the overall $K^+K^-$ yield in elementary p+p
collisions\cite{anke_1}. In this view, the $K^-$ production in
heavy-ion collisions at SIS energies can not be explained
exclusively by the strangeness exchange mechanism and the
processes mentioned above must also be taken into account to
achieve a complete description.
\section{Comparison to statistical models}
\begin{figure}[htb]
\begin{minipage}[t]{75mm}
      \begin{picture}(200,180)(0,0)
      \put(-32,-2.6){ \includegraphics[width=1.1\textwidth,height=.915\textwidth]
          {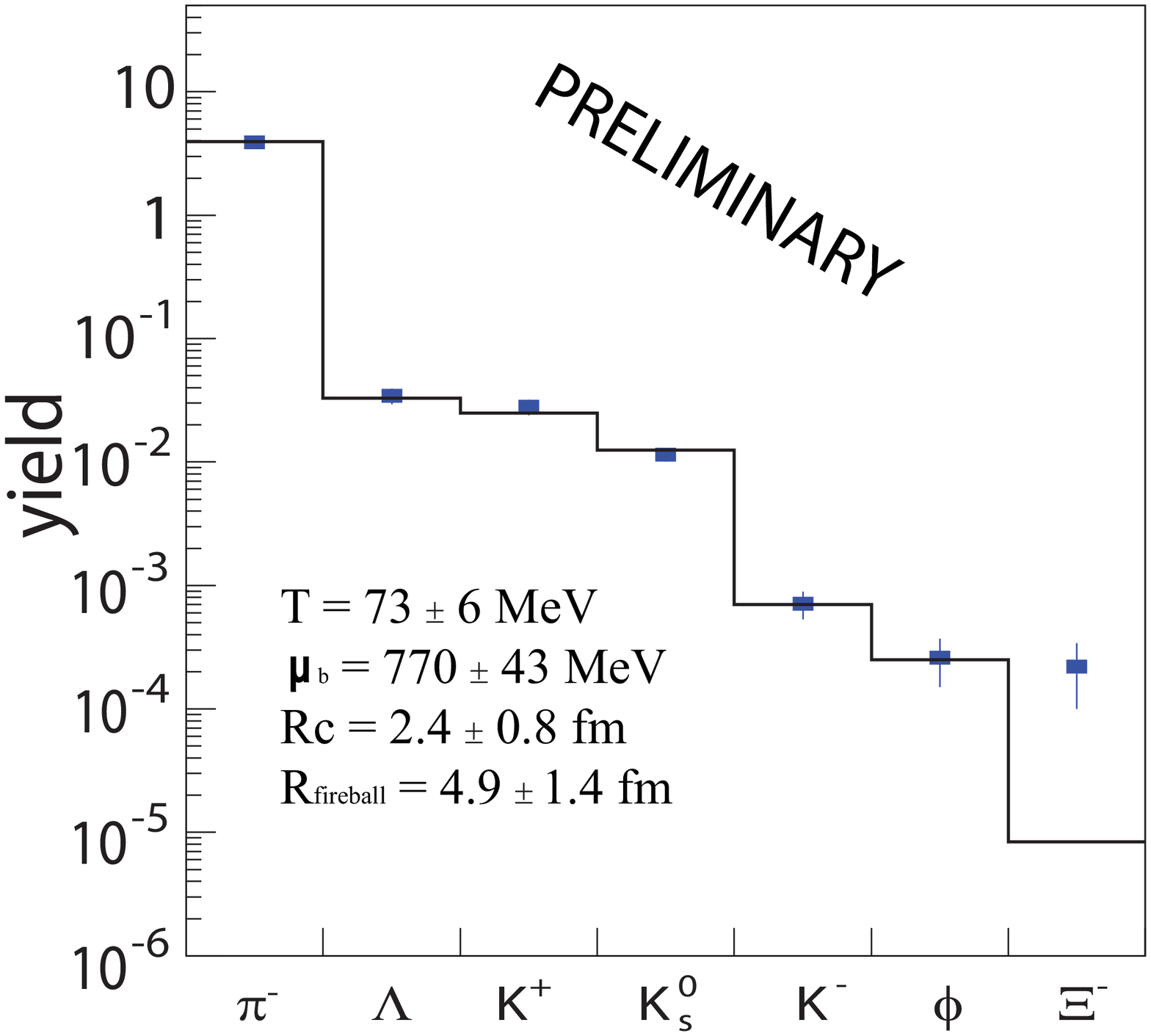}}
      \end{picture}
\caption{Hadron multiplicities (extrapolated to full solid angle)
measured by HADES in Ar+KCl collisions at 1.756 AGeV compared to
the THERMUS model\cite{thermus_1} calculations.}
\label{fig:thermus}
\end{minipage}
 \hspace{\fill}
\begin{minipage}[t]{75mm}
      \begin{picture}(200,180)(0,0)
      \put(0,-15){ \includegraphics[width=.97\textwidth,height=.97\textwidth]
          {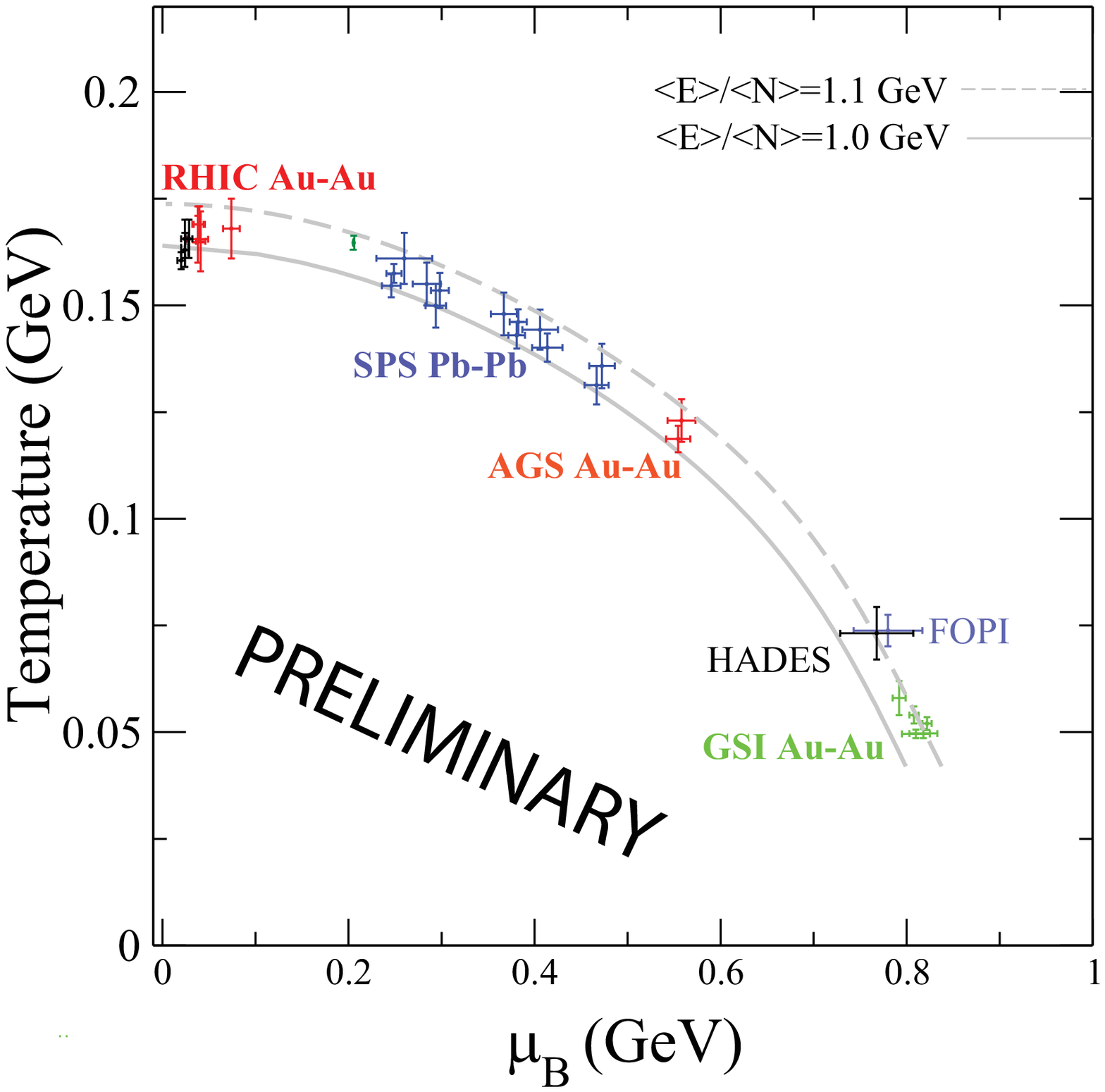}}
      \end{picture}
\caption{Chemical freeze-out parameters obtained in the
statistical thermal model (for details see \cite{cleymans_1}). The
HADES point corresponds to Ar+KCl collisions at 1.756 AGeV.}
\label{fig:statmodel}
\end{minipage}
\end{figure}

The yields of reconstructed hadron species have been extrapolated
to the full solid angle and compared to the result from a fit with
the statistical model THERMUS\cite{thermus_1} as shown in Fig.
\ref{fig:thermus}. The measured yields nicely agree with the
results of the model, except for the $\Xi^-$. One should note that
in this approach a good description of the $\phi$ meson yield is
obtained without assuming any strangeness suppression (net
strangeness content of the $\phi$ is S=0). This is very different
as compared to higher energies where $\phi$ meson does not behave
as a strangeness neutral object but rather as an object
with net strangeness between 1 and 2 \cite{strgns_phi}.\\

The $\Xi^-$ baryon yields measured in heavy-ion collisions above
the production threshold at RHIC\cite{rhic_1}, SPS\cite{sps_1} and
AGS\cite{ags_1} nicely agree with the statistical model
predictions. On the contrary the result of the first $\Xi^-$
measurement below the production threshold published by
HADES\cite{hades_ksi}, shows a deviation of about an order of
magnitude from the calculations (Fig. \ref{fig:thermus}).
 Using the measured hadron multiplicities the statistical model predicts
that the chemical freeze-out of the Ar+KCl collision at 1.765 AGeV
occurs at a temperature of T=73$\pm$6MeV and at chemical baryon
potential of $\mu$=770$\pm$43MeV. The strangeness correlation
radius\cite{hades_kaons} of R$_c=$2.4$\pm$0.8fm was used which is
significantly smaller than the radius of fireball
R$_{fireball}$=4.9$\pm$1.4fm.\\ This result nicely follows the
striking regularity shown by particle yields at all beam energies
\cite{cleymans_1}, as presented in Fig. \ref{fig:statmodel}. For
all available energies, starting from the highest at RHIC down to
the lowest at SIS, the measured particle multiplicities are
consistent with the assumption of chemical equilibrium which sets
in at the end of the collision phase. Only two parameters (the
temperature and baryon chemical potential) are needed within a
thermal-statistical model to describe particle yields in a very
systematic way at a given collision energy \cite{cleymans_1}. As
one can also see, all experimental results are in good agreement
with a fixed-energy-per-particle condition
$<E>$$/$$<N>$$\approx1GeV$, which is one of the available
freeze-out criteria\cite{cleymans_1}.

The new HADES results on strangeness production shed new light on
the understanding of kaon production mechanisms in HI collisions,
namely the results have provided compelling evidence that the
contribution from the $\phi$ decay to the $K^-$ yield has to be
also taken into account. The measured hadron yields have been
found in general to be in good agreement with statistical model
predictions, besides the $\Xi^-$, which is produced far below the
production threshold and shows a considerable deviation from the
statistical model. The already performed experiments p+p at 3.5
GeV and p+Nb at 3.5 GeV and further planned HADES experiments with
heavier systems, like Au+Au will deliver new valuable data on
strangeness production. The on-going upgrade of the HADES
spectrometer will increase its performance and capability and the
installed Forward Wall detector will allow for reaction plane
reconstruction in all upcoming runs, allowing to study kaon flow
observables as well.

\end{document}